\documentclass{ws-procs9x6}

\def\beqn{\begin{eqnarray}}
\def\eeqn{\end{eqnarray}}
\def\be{\begin{equation}}
\def\ee{\end{equation}}

\begin{document}

\title{Resonance Structure and Polarizability of the nucleon}

\author{D. Drechsel}

\address{Institut f\"ur Kernphysik, Johannes Gutenberg-Universit\"at, \\
Becherweg 45, \\
55099 Mainz, Germany\\
E-mail: drechsel@kph.uni-mainz.de}


\maketitle

\abstracts{ The main features of the resonance structure of the nucleon
are discussed, particular with regard to the helicity dependence of
real and virtual photoabsorption. The dependence of the partial cross
sections on both the resonance helicity amplitudes and the
electromagnetic multipoles is outlined. The general structure of the
Compton tensor is reviewed and applied to the special cases of real to
real, virtual to real, and virtual to virtual Compton scattering.
Recent theoretical developments in dispersion relations are presented,
together with a short overview regarding static, dynamical, and
generalized polarizabilities of the nucleon as well as the status of
the Gerasimov-Drell-Hearn sum rule and related integrals.
}

\section{The Resonance Structure of the Nucleon}

On top of a nonresonant background, the total photoabsorption cross
section of the nucleon exhibits three resonance regions, mainly
corresponding to the resonances $\Delta(1232)$, N$^*$(1520), and
N$^*$(1680) with dominance of magnetic dipole (M1), electric dipole
(E1), and electric quadrupole (E2) radiation,
respectively~\cite{Bab98}. Above the third resonance region, the
absorption cross section levels off at a value of about 120~$\mu b$. At
the highest measured energies, at total $c.m.$ energy $W\simeq200$~GeV,
the cross section increases slowly~\cite{DESY}, in accordance with the
soft pomeron exchange mechanism.

\begin{figure}[th]
\vspace{0cm}
\epsfxsize=12.5cm \epsfysize=5cm
\centerline{\epsffile{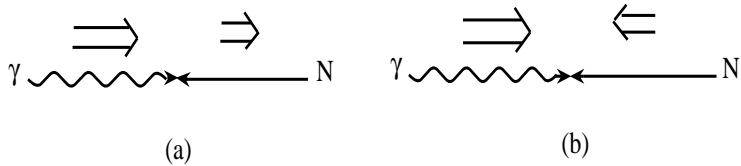}} \vspace{-1.5cm}
\caption{Spin and helicity of a double polarization experiment.
The arrows $\Longrightarrow$ denote the spin projections on the
photon momentum, the arrows $\longrightarrow$ the momenta of the
particles. The spin projection and helicity of the photon is
assumed to be $h_{\gamma}=1$. The spin projection and helicity of the
target nucleon $N$ are denoted by $S_z$ and $h$, respectively, and
the eigenvalues of the excited system $N^{\ast}$ by the
corresponding primed quantities. \newline a) Helicity 3/2:
Transition $N^{\ast}(S_z=1/2,\ h_N=-1/2)\rightarrow
N^{\ast}(S_z=h_{N^{\ast}}=3/2)$, which changes the helicity by 2 units.
\newline
b) Helicity 1/2: Transition $N(S_z=-1/2,\
h_N=+1/2)\rightarrow(S_z=h_{N^{\ast}}=+1/2)$, which conserves the helicity.
\label{DDfig2.1}}
\end{figure}

The resonance structures stick out
much more clearly in the recent double-polarization experiments
performed at MAMI~\cite{Ahr00} and ELSA~\cite{Hel02}. In these
investigations, circularly polarized photons are absorbed by protons
with polarization parallel and antiparallel to the photon momentum,
which leads to hadronic excitations with helicities $3/2$ and $1/2$,
respectively, as described in Fig.~\ref{DDfig2.1}.
The difference of the two helicity cross sections,
$\sigma_{3/2}-\sigma_{1/2}$, displays the following features as
function of the photon $lab$ energy $E_{\gamma}$: (I) Negative
values near threshold due to S-wave pion production, (II) a large
(positive) peak at the position of the $\Delta(1232)$, (III)
another peak somewhat below the N$^*(1520)$, mostly due to the
onset of two-pion production, (IV) a somewhat smaller peak in the
third resonance region, and (V) very small helicity differences at
the higher energies, with a possible cross-over to negative values
above $E_{\gamma}\approx2$~GeV. In other words, the large
background showing in the total cross section is ``helicity
blind'' and drops out in measurements of the helicity difference.

By use of inelastic electron scattering the absorption cross
section may be generalized to virtual photons characterized by
energy transfer $\nu$ and four-momentum squared, $Q^2$, and expressed
in terms of a virtual photon flux factor $\Gamma_V$ and four partial cross
sections~\cite{Dre01},
\beqn
\frac{d\sigma}{d\Omega\ dE'} = \Gamma_V \sigma
(\nu,Q^2)\ , \label{DDeq2.1a}
\eeqn
\beqn
\sigma =
\sigma_T+\epsilon\sigma_L-hP_x\sqrt{2\epsilon(1-\epsilon)}\
         \sigma_{LT}-hP_z\sqrt{1-\epsilon^2}\ \sigma_{TT}\ ,
\label{DDeq2.1b}
\eeqn
with the photon polarization $\epsilon$.
The four partial cross sections $\sigma_T,\ \sigma_L,\
\sigma_{TT}$ and $\sigma_{LT}$ are uniquely determined by the
quark structure functions, $F_1,\ F_2,\ g_1$ and $g_2$.
Furthermore, the transverse cross sections are related to the
partial cross sections $\sigma_{3/2}$ and  $\sigma_{1/2}$ of
Fig.~\ref{DDfig2.1}(a) and~\ref{DDfig2.1}(b), respectively,
$\sigma_T = (\sigma_{1/2}+\sigma_{3/2})/2,\
\sigma_{TT} = (\sigma_{1/2}-\sigma_{3/2})/2$.

While the partial cross sections are usually considered as
functions of $\nu$ and $Q^2$, the quark structure functions are
written as functions of $x$ and $Q^2$, where $x=Q^2/2m\nu$ is the
Bj\"orken scaling variable. In the region of deep inelastic
scattering (DIS), where the masses of the constituents can be neglected,
the observables should only depend on the specific
ratio given by Bj\"orken's definition.

On the hadronic side the resonances can be described by either
their helicity amplitudes or the multipoles of the pion-nucleon
system in the final state. The helicity $h$ of a particle is
given by the projection of its spin $\vec{s}$ onto the direction
of its momentum, $\hat{k}$. Since the projection of orbital
momentum onto this axis vanishes, only the intrinsic spin is
involved, and consequently the nucleon takes the values
$h_N=\pm\frac{1}{2}$, while the virtual photon has
$h_{\gamma}=\pm1$ and 0. The helicity is a pseudoscalar, invariant
under rotations but changing sign under the parity transformation.
If the hadronic states have good parity, only 3 of the 6
helicity amplitudes are independent: the two transverse amplitudes
corresponding to Fig.~\ref{DDfig2.1}, \newline
$A_{3/2}(h_N=-\frac{1}{2},\ h_{\gamma}=1\rightarrow
h_{N^*}=\frac{3}{2})$,
$A_{1/2}(h_N=\frac{1}{2},\ h_{\gamma}=1\rightarrow
h_{N^*}=\frac{1}{2})$, and the
longitudinal amplitude
$S_{1/2}(h_N=-\frac{1}{2},\ h_{\gamma}=0\rightarrow
h_{N^*}=\frac{1}{2})$.

In the case of an isolated resonance, the partial cross sections
are related to the helicity amplitudes as follows:
\be
\sigma_{1/2}\sim|A_{1/2}|^2,\ \sigma_{3/2}\sim|A_{3/2}|^2,\
\sigma_{L}\sim|S_{1/2}|^2,\ \sigma_{LT}\sim S^*_{1/2}A_{1/2}\ .
 \label{DDeq2.5}
\ee
The alternative description is in terms of electric $(E)$, magnetic
$(M)$, and ``scalar'' $(S)$ multipoles. These are further
characterized by the relative orbital momentum $l$ of the
pion-nucleon final state and a sign, which is plus or minus if the
total angular momentum $J$ is equal to $l+\frac{1}{2}$ or
$l-\frac{1}{2}$, respectively. The multipole content of the most
important resonances is given in Table~\ref{mult_contr}. As an
example the $P_{33}$ partial wave corresponds to the
$\Delta(1232)$ with spin and isospin $3/2$, orbital momentum
$l=1$,
and positive parity. Its multipoles are therefore $M_{1+}$
(magnetic dipole radiation), $E_{1+}$ (electric quadrupole), and
$S_{1+}$ (Coulomb quadrupole).

\begin{table}[htb]
\tbl{The multipole contributions to the partial cross sections for
some selected resonances. The entries have to be multiplied by
overall kinematical factors. } {\footnotesize
\begin{tabular}{|c|l|l|l|l|}
\hline
resonance & $\sigma_L$ & $\sigma_{3/2}$ & $\sigma_{1/2}$ & $\sigma_{LT}$\\
\hline $P_{33}$ & $8|S_{1+}|^2$ & $3|E_{1+}-M_{1+}|^2$ &
$|3E_{1+}+M_{1+}|^2$ &
$+2S_{1+}^{\ast}(3E_{1+}+M_{1+})$\\
$P_{11}$ & $|S_{1-}|^2$ & $-$ & $2|M_{1-}|^2$ & $-S_{1-}^{\ast} M_{1-}$ \\
$D_{13}$ & $8|S_{2-}|^2$ & $3|E_{2-}+M_{2-}|^2$ &
$|E_{2-}-3M_{2-}|^2$ &
$+2S_{2-}^{\ast} (E_{2-}-3M_{2-})$ \\
$S_{11}$ & $|S_{0+}|^2$ & $-$ & $2|E_{0+}|^2$ & $+S_{0+}^{\ast}
E_{0+}$ \\
$D_{15}$ & $27|S_{2+}|^2$ & $12|E_{2+}-M_{2+}|^2$ &
$6|2E_{2+}+M_{2+}|^2$ & $+9S_{2+}^{\ast} (2E_{2+}+M_{2+})$ \\
$F_{15}$ & $27|S_{3-}|^2$ & $12|E_{3-}+M_{3-}|^2$ &
$6|E_{3-}-2M_{3-}|^2$ & $+9S_{3-}^{\ast} (E_{3-}-2M_{3-})$ \\
\hline
\end{tabular}\label{mult_contr} }
\end{table}

The helicity structure of the resonance region changes with
increasing ``virtuality'' $Q^2$, as can be seen from
Fig.~2 of Ref.~\cite{Dre01}. The main features are: (I)~The $\sigma_{1/2}$
contribution of pion threshold production decreases rapidly with
increasing $Q^2$, (II)~the $\sigma_{3/2}$ dominated $\Delta(1232)$
cross section remains nearly constant up to
$Q^2\approx0.5$~GeV$^2$ but drops strongly thereafter, (III)~the
helicity difference $\sigma_{3/2}-\sigma_{1/2}$ in the
second and third resonance regions of the proton is positive for $Q^2=0$
but becomes negative already at
$Q^2\approx0.5$~GeV$^2$. This feature can be
understood by strong electric multipoles $E_{2-}$ and
$E_{3-}$ for real photons but an increase of the
corresponding magnetic multipoles with virtuality $Q^2$ (see
Table~\ref{mult_contr}), such that the helicity amplitude $A_{3/2}$
dominates at $Q^2=0$ while $A_{1/2}$ takes over for large $Q^2$,
because it conserves the helicity.

\section{Compton Scattering}

 By use of Lorentz and gauge invariance, crossing symmetry,
parity and time reversal invariance, the amplitude for Compton
scattering takes the form~\cite{Lvo97}
\be
   \left\langle \chi_f\mid{\mathcal T}\mid\chi_i\right\rangle=
   \epsilon'^{\ast}_{\mu}\epsilon_{\nu}
   \sum_{\lambda=1}^{\lambda_{\mbox{\scriptsize{max}}}}
   \left\langle \chi_f\mid{\mathcal O}_{\lambda}^{\mu\nu}\mid\chi_i
   \right\rangle \tilde{A}_{\lambda}(s,t)\ ,
 \label{DDeq3.1}
\ee
where ${\mathcal O}_{\lambda}^{\mu\nu}$ are Lorentz tensors constructed
from kinematical variables and $\gamma$ matrices, and
$\tilde{A}_{\lambda}$ are Lorentz scalars. In the $c.m.$ frame, these
Lorentz structures can be reduced to Pauli matrices $\vec{\sigma}$
combined with unit vectors in the directions of the initial $(\hat{k})$
and final $(\hat{k}')$ photons. The polarization vectors of the initial
and final photons are denoted by $\vec{\epsilon}$ and $\vec{\epsilon}\ '$
for transverse polarization, and $\hat{k}$ and $\hat{k}'$ for
longitudinal polarization. Special cases are:

\vspace{0.3cm}
\noindent
{\it I. Real Compton Scattering (RCS)}

\noindent
From helicity arguments there should be $2^4=16$ amplitudes, but
parity conservation reduces this number to 8. In the {\em c.m.}
frame the transition operator can be cast into the form
\begin{eqnarray}
\label{DDeq3.2}
 {\mathcal T}_{RR}&=&A_1(\omega , \theta)\vec{\epsilon}\ '^{\ast}
    \cdot \vec{\epsilon}+
    A_2(\omega , \theta)\vec{\epsilon}\ {'}^{\ast} \cdot \hat{k}\vec{\epsilon}
        \cdot \hat{k}{'}\nonumber \\
  && +  iA_3(\omega , \theta)\vec{\sigma} \cdot
        (\vec{\epsilon}\ {'}^{\ast} \times \vec{\epsilon})+
      iA_4(\omega , \theta)\vec{\sigma} \cdot (\hat{k}{'} \times \hat{k})
        \vec{\epsilon}\ {'}^{\ast} \cdot \vec{\epsilon} \nonumber \\
  && +  iA_5(\omega , \theta)\vec{\sigma} \cdot \left[(\vec{\epsilon}\ {'}^{\ast}
       \times \hat{k})\vec{\epsilon} \cdot \hat{k}{'}-(\vec{\epsilon}
       \times \hat{k}')\vec{\epsilon}\ {'}^{\ast}\cdot \hat{k} \right ]  \nonumber \\
  && +  iA_6(\omega , \theta)\vec{\sigma} \cdot \left[(\vec{\epsilon}\ {'}^{\ast}
       \times \hat{k}')\vec{\epsilon} \cdot \hat{k}{'}-(\vec{\epsilon}
       \times \hat{k})\vec{\epsilon}\ {'}^{\ast}\cdot \hat{k} \right ] \nonumber \\
  && +  iA_7(\omega , \theta)\vec{\sigma} \cdot \left[(\vec{\epsilon}\ {'}^{\ast}
       \times \hat{k})\vec{\epsilon} \cdot \hat{k}{'}+(\vec{\epsilon}
       \times \hat{k}')\vec{\epsilon}\ {'}^{\ast}\cdot \hat{k} \right ]  \nonumber \\
  && +  iA_8(\omega , \theta)\vec{\sigma} \cdot \left[(\vec{\epsilon}\ {'}^{\ast}
       \times \hat{k}')\vec{\epsilon} \cdot \hat{k}{'}+(\vec{\epsilon}
       \times \hat{k})\vec{\epsilon}\ {'}^{\ast}\cdot \hat{k}
       \right ]\ ,
\end{eqnarray}
where $\omega$ is the energy and $\theta$ the scattering angle of
the photon in the {\it c.m.} system.

By use of time reversal $(\hat{k}\leftrightarrow-\hat{k}',\
\vec{\epsilon}\leftrightarrow\vec{\epsilon}\ {'}^{\ast},\
i\vec{\sigma}\rightarrow-i\vec{\sigma})$ $A_7$ and
$A_8$ change sign, and therefore
vanish if time reversal invariance holds. Hence the
transition operator for RCS is described by the 2 scalar
amplitudes $A_1$ and $A_2$, and 4 spin amplitudes $A_3$ to $A_6$.

Due to the transversality condition $\vec{\epsilon}\cdot\hat{k}=
\vec{\epsilon}\ {'}\cdot\hat{k}'=0$, only 2 amplitudes contribute
in the forward direction, $\hat{k}=\hat{k}'$, and with some
change of notation, $f=A_1$ and $g=A_3$, the forward scattering amplitude
reads
\be
  {\mathcal T} (\omega, \theta = 0) = \hat{\epsilon}\,'^{\ast} \cdot \hat{\epsilon}
  \ f(\omega) + i (\hat{\epsilon}\,'^{\ast} \times \hat{\epsilon}) \cdot
  \vec{\sigma}\ g(\omega)\ .
\label{DDeq3.3}
\ee
%

%\vspace{0.5cm}
\noindent
{\it II. Virtual Compton Scattering (VCS)}

\noindent Since this process describes a reaction rather than a
scattering, namely the transition from a virtual photon $\gamma^{\ast}$
with $k^2=-Q^2<0$ to a real, massless photon $\gamma$, time reversal
does not provide any constraint, and all 8 combinations of
Eq.~(\ref{DDeq3.2}) are allowed. In addition there are 4 interference
terms between the longitudinal helicity of the virtual photon and the
transverse helicity of the real photon, which can be found by replacing
the transverse polarization vector $\vec{\epsilon}$ in
Eq.~(\ref{DDeq3.2}) by the longitudinal polarization vector $\hat{k}$.
The VCS Compton tensor then takes the form
\begin{eqnarray}
\label{DDeq3.4}
{\mathcal T}_{VCS}&=& {\mathcal T}_{RR}+ \vec{\epsilon}\ {'}^{\ast}\cdot\hat{k}\, A_9
     + i\vec{\sigma}\cdot (\hat{k}'\times\hat{k})\,
       \vec{\epsilon}\ {'}^{\ast}\cdot\hat{k}\, A_{10} \nonumber \\
     && + i\vec{\sigma}\cdot (\vec{\epsilon}\ {'}^{\ast}\times\hat{k})
     \, A_{11}
     + i\vec{\sigma}\cdot (\vec{\epsilon}\ {'}^{\ast}\times\hat{k}')
     \, A_{12} \ .
\end{eqnarray}
Altogether there appear 3 scalar and 9 spin amplitudes in VCS.

\vspace{0.3cm}
\noindent
{\it III. Doubly virtual Compton scattering (VVCS)}

\noindent In comparing with the case of VCS, we find 4 more
longitudinal-transverse interference terms by replacing
$\vec{\epsilon}\ {'}^{\ast}\rightarrow \vec{\epsilon}$ and
$\hat{k}\rightarrow\hat{k}'$ in the last 4 terms of
Eq.~(\ref{DDeq3.4}). Furthermore, there appear 2 terms constructed from
the longitudinal-longitudinal combination (Replace $\vec{\epsilon}\
{'}^{\ast}$ by $\hat{k}'$ in the terms with $A_9$ and $A_{11}$ of
Eq.~(\ref{DDeq3.4})!). The complete form is
\begin{eqnarray}
\label{DDeq3.5}
{\mathcal T}_{VVCS} &=& {\mathcal T}_{VCS} + \vec{\epsilon}\cdot\hat{k}'\,A_{13} +
  i\vec{\sigma}\cdot(\hat{k}\times\hat{k}')\vec{\epsilon}
  \cdot\hat{k}'\,A_{14}
   + i\vec{\sigma}\cdot(\vec{\epsilon}
  \times\hat{k}')\,A_{15}\nonumber \\ && + i\vec{\sigma}\cdot(\vec{\epsilon}
  \times\hat{k})\, A_{16}
   + \hat{k}\cdot\hat{k}'\, A_{17} +i\vec{\sigma}\cdot
  (\hat{k}'\times\hat{k})\, A_{18} \ .
\end{eqnarray}
Up to this point we have considered the general reaction
$\gamma^{\ast}(Q^2)\rightarrow\gamma^{\ast}(Q{}'^2)$. If we
restrict the discussion to photons with equal virtuality in the
initial and final states, $Q^2=Q{}'^2$, time reversal invariance
leads to additional constraints: $A_7=A_8=0$ as for RCS, and the
relations
$A_9=A_{13}$,\ $A_{10}=-A_{14}$,\ $A_{11}=A_{15}$,
\ $A_{12}=A_{16}$, which leaves 12 independent amplitudes. In the
forward direction the amplitude reads
\begin{eqnarray}
\label{DDeq3.6}
 {\mathcal T}_{VVCS} (\theta  =  0)
& = & \hat{\epsilon}\,'^{\ast} \cdot \hat{\epsilon}
  \ f_T(\omega ,Q^2) + f_L(\omega,Q^2)
  + i (\hat{\epsilon}\,'^{\ast}
  \times \hat{\epsilon}) \cdot
  \vec{\sigma}\ g_{TT}(\omega ,Q^2) \nonumber
 \\
 &&
  +i\vec{\sigma}\cdot
  \left[(\vec{\epsilon}-\vec{\epsilon}\ {'}^{\ast})\times
  \hat{k}\right]g_{LT}(\omega,Q^2)\ ,
\end{eqnarray}
with $f_T=A_1,\ f_L=A_{17},\ g_{TT}=A_3,\
g_{LT}=A_{11}+A_{12}+A_{15}+A_{16}$. The 4 amplitudes
$\{f_T,f_L,g_{TT},g_{LT}\}$ can be constructed from the 4 inclusive
electroproduction cross sections
$\{\sigma_T,\sigma_L,\sigma_{TT},\sigma_{LT}\}$ by means of
forward dispersion relations.

\subsection{Real Compton Scattering (RCS)}

The forward amplitude for RCS, Eq.~(\ref{DDeq3.3}), can be
determined by the double polarization experiment of
Fig.~\ref{DDfig2.1}. The crossing symmetry requires that $f$ is an
even and $g$ an odd function of the photon $lab$ energy
$\nu=E_{\gamma}$. These amplitudes can be constructed by
dispersion relations (DR) based on analyticity (required by causality)
and unitarity (the optical theorem in forward direction),
\beqn
\label{DDeq3.7}
{\mbox{Re}}\ f(\nu) =
f(0)+\frac{\nu^2}{2\pi^2}\,
\int_{\nu_0}^{\infty}\,\frac{\sigma_T(\nu')}
{\nu'^2-\nu^2}\,d\nu'\ ,
\eeqn

\beqn
\label{DDeq3.8}
{\mbox{Re}}\ g(\nu) = \frac{\nu}{4\pi^2}\,{\mathcal{P}}\,
\int_{\nu_0}^{\infty}\,\frac{\sigma_{1/2}(\nu')-\sigma_{3/2}(\nu')}
{\nu'^2-\nu^2}\,\nu'd\nu'\ .
\eeqn

These results may be expanded into a Taylor series for small values
of $\nu$ and compared to the low energy theorem of Low~\cite{Low54},
and Gell-Mann and Goldberger~\cite{Gel54}, which expresses the leading
term of the amplitudes by the charge $e_N$ and the anomalous magnetic
moment $\kappa_N$,
\begin{eqnarray}
f(\nu) & = & -\frac{e^2\,e_N^2}{4\pi M} + (\alpha+\beta)
\,\nu^2+ {\mathcal{O}}(\nu^4) \ , \label{DDeq3.9} \\
g(\nu) & = & -\frac{e^2\kappa^2_N}{8\pi M^2}\,\nu +
\gamma_0\nu^3 + {\mathcal{O}}(\nu^5) \ . \label{DDeq3.10}
\end{eqnarray}

The result of the comparison is Baldin's sum
rule~\cite{Bal60},
\beqn
\label{DDeq3.11}
\alpha + \beta = \frac{1}{2\pi^2}\,
\int_{\nu_0}^{\infty}\,\frac{\sigma_T(\nu')}{\nu'^2}
\,d\nu'\ ,
\eeqn
the sum rule of Gerasimov~\cite{Ger65}, Drell and
Hearn~\cite{Dre65},
\beqn
\label{DDeq3.12}
\frac{\pi e^2\kappa^2_N}{2M^2}=
\int_{\nu_0}^{\infty}\,\frac{\sigma_{3/2}(\nu')
-\sigma_{1/2}(\nu')}{\nu'}\,d\nu' \, \equiv I \, ,
\eeqn
and the sum rule for the forward spin
polarizability~\cite{Gel54,GGT54},
\beqn
\label{DDeq3.13}
\gamma_0= \,-\,\frac{1}{4\pi^2}\,\int_{\nu_0}^{\infty}\,
\frac{\sigma_{3/2}(\nu')-\sigma_{1/2}(\nu')}
{\nu'^3}\,d\nu'\ .
\eeqn

\begin{figure}[ht]
\vspace{-.25cm} \epsfxsize=11cm
\centerline{\epsffile{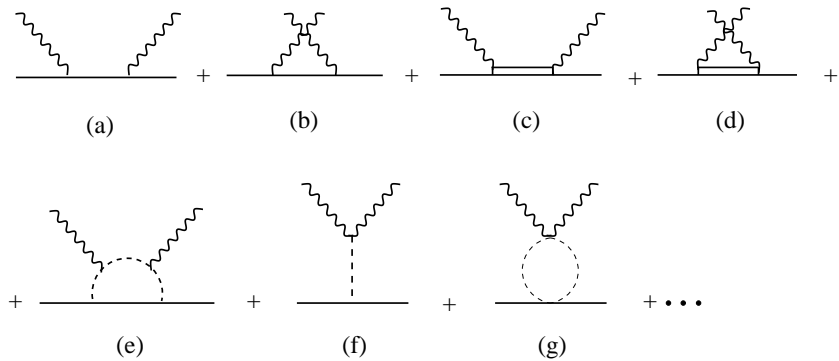}} \caption[]{Graphs
contributing to Compton scattering off the nucleon. Upper row: The
direct (a) and crossed (b) Born diagrams with intermediate
nucleons, a typical resonance excitation in the $s$-channel (c), and
its crossed version (d). Lower row: Typical mesonic contributions
with photon scattering off an intermediate pion (e), the pion pole
diagram (f), and a correlated two-pion exchange such
as the ``$\sigma$ meson'' (g).} %\label{DDfig2.2.5} }
\label{fig:graphs_rcs}
\end{figure}

Real Compton scattering at general angles requires to set up DR for all
6 amplitudes, i.e., to evaluate the real part of these amplitudes by the
pole term contributions (see graphs a, b, and f of
Fig.~\ref{fig:graphs_rcs})
and integrals over the imaginary part. The
latter can be constructed from the respective photoproduction
processes, e.g., $\gamma+N\rightarrow\pi+N$ and production of heavier
systems like $\pi\pi,\eta,$ etc., as described by the SAID or MAID
analyses.
For example, Fig.~\ref{fig:graphs_rcs}(c) represents a resonance
contribution yielding
imaginary parts in the s-channel region, (g) leads to imaginary parts
in the t-channel region, and the pion-loop diagram (e) may produce
imaginary parts in both s and t channel, depending on how one ``cuts''
the diagram.

Our recent work~\cite{Dre02} includes 4 types of DR: DR at constant $t$ and
hyperbolic DR, in both an unsubtracted and a subtracted version, while
the previous analysis was essentially based on unsubtracted DR at
$t=const.$ The dispersion integrals run over the variable $\nu$, which
for general kinematics is defined as the average over the initial and
final photon $lab$ energies, $\nu=(E_{\gamma} +  E'_{\gamma})/2$,
typically up to $E_{\gamma}\approx1.5$~GeV. In the case of unsubtracted DR,
some of these integrals do not converge, and the contribution from the
higher energies is modelled by t-channel poles, in particular the
(known) $\pi^{\circ}$ pole and the exchange of a ``$\sigma$ meson''
with spin and isospin zero. In order to reduce the phenomenology
involved with heavier intermediate mass states and the $\sigma$ meson,
we have subtracted such DR at $\nu=0$. The subtraction function is then
constructed by once subtracted dispersion integrals in $t$, by use of
experimental information on the reaction $\gamma+\gamma\rightarrow
\pi+\pi$ and an extrapolation of pion-nucleon scattering into
the unphysical region~\cite{DGP99}. While the subtraction improves the convergence
at large $\nu$, it has a shortcoming at small $\nu$: The
dispersion integrals get contributions from outside of the physical
region, for which the integrand is constructed by an extension of the
 partial wave
expansion to unphysical angles. This limits the calculation to low
partial waves or photon energies below the $\Delta$ resonance.

In order to improve the convergence for larger values of $t$,
fixed-angle (or hyperbolic) DR have been proposed~\cite{Ber74} and
applied to Compton scattering~\cite{Hol94,Lvo99}. In particular for
$\theta_\mathrm{lab}=180^{\circ},$ the path of integration runs along
the lower boundary of the $s$-channel region from infinity to the origin of the
Mandelstam plane (``$s$-channel contribution''),
and then continous along a path in the upper half-plane
(``$t$-channel contribution''). The recent calculations of
Pasquini~\cite{Pas02} explain, for the first time, the puzzle of the
large difference of the electric $(\alpha)$ and magnetic $(\beta)$
polarizabilities known from backward Compton scattering. For example,
the recent experiment with TAPS at MAMI resulted in
$(\alpha-\beta)_{exp}=[10.7\pm0.6{\mbox{(stat)}}\pm0.8{\mbox{(syst)}}]$,
here and in the following in units of $10^{-4}$~fm$^3$, while DR at
$t=const$ predict $\alpha-\beta\approx-6$ for the $\pi N$ intermediate
states and an upper limit of integration $E_{\gamma}=1.5$~GeV. Within
the framework of unsubtracted DR at $t=const$, the discrepancy is then
described by heavier intermediate states and the contribution of a
$\sigma$-meson pole in the t-channel. However, within the framework of
hyperbolic DR~\cite{Pas02}, the result is
$(\alpha-\beta)_{hyp}=10.9$, obtained by adding the s-channel
contribution at $\theta=180^{\circ}$, $(\alpha-\beta)_s=-5.6$ and its
continuation into the t-channel, $(\alpha-\beta)_t=16.5$.

Besides the static polarizabilities, it is also possible to define
``dynamical'' polarizabilities as function of $\nu$, e.g., the
dynamical electric dipole polarizability $\alpha_{E1}(\nu)$ with
$\alpha_{E1}(0)=\alpha$. This procedure requires a decomposition of the
Compton amplitudes into a partial wave series~\cite{Bab98a} of dipole
and higher order multipoles including retardation or dispersion
effects. The dynamical polarizabilities allow for a very detailed study
of the internal degrees of freedom. For example, $\alpha_{E1}$ and
$\alpha_{E2}$ clearly show cusp effects at the pion
threshold, and $\beta_{M1}$ exhibits
 the $\Delta$-resonance structure, with its real part passing through
 zero at the resonance position.
Except for $\beta_{M2}$, the HBChPT calculation nicely reproduces the
 results of DR~\cite{Pas02}.

\subsection{Virtual Compton Scattering (VCS)}

The DR formalism has been extended to VCS, a tool to extract
generalized polarizabilities (GPs) by means of radiative electron
scattering. These GPs are functions of the virtuality of the incident
photon and describe, in some sense, the spatial distribution of the
polarizabilities. The first unpolarized VCS observables have been
obtained from MAMI~\cite{Roc00} at a virtuality $Q^2$ = 0.33 GeV$^2$,
and recently at JLab \cite{Fon02} at higher virtualities, 1~GeV$^2< Q^2
< 2~$GeV$^2$. Further experimental programs are underway at the
intermediate energy electron accelerators (MIT-Bates \cite{Mis97}, MAMI
\cite{dHo01}, and JLab~\cite{Hyd02}) to measure both unpolarized and
polarized VCS observables. The existing data indicate a $Q^2$
dependence of the electric GP similar to a dipole form factor, whereas
the magnetic GP follows a more complicated $Q^2$ behavior. As was
already shown for RCS, the magnetic dipole transition involves a strong
cancellation between the diamagnetism due to the pion cloud effects
(essentially the ``asymptotic'' or t-channel contribution) and the
paramagnetism due to resonance excitation (essentially the quark
spin-flip transition to the $\Delta$
resonance in the s-channel). Since the cloud effects have a considerably
longer range in
space than the resonance structures, the $Q^2$ behavior of the magnetic
GP is able to disentangle both physical mechanisms, which is already
displayed in the existing data. Given this initial success, future
experiments to measure VCS observables in the $\Delta$ region hold the
promise to extract GPs with an enhanced precision, within the DR
formalism.

\subsection{Doubly Virtual Compton Scattering}

As has been shown in Eq.~(\ref{DDeq3.6}), forward VVCS is described by
4 independent amplitudes, which can be constructed from the partial
cross sections of Eq.~(\ref{DDeq2.1b}). In this sense the sum rules of
Eqs.~(\ref{DDeq3.11}) to (\ref{DDeq3.13}) may be generalized to
virtual photons~\cite{Dre02,Hem02}, and 2 further sum rules can be
constructed for the amplitudes $f_L$ and $g_{LT}$ involving the
longitudinal photon. In particular the ``generalized GDH integral''
$I(Q^2)$ was recently measured at the Jefferson Lab~\cite{deJ02} for both the
proton and the neutron. These data indicate
a dramatic ``phase transition'' between the resonance dominated region
at $Q^2\lesssim1$~GeV$^2$ and the regime of DIS at the larger
values of $Q^2$, in qualitative agreement with the results of DR at low
$Q^2$ and perturbative QCD at large $Q^2$~\cite{Ans02}.

Due to the weighting with the energy denominators, the generalized
forward spin polarizability $\gamma_0(Q^2)$ and the
longitudinal-transverse polarizability $\delta_{LT}(Q^2)$, related to the amplitudes
$g_{TT}$ and $g_{LT}$ of Eq.~(\ref{DDeq3.6}), respectively, are
determined by resonance and pionic degrees of
freedom. The observable $\gamma_0(Q^2)$ was recently determined at
$Q^2=0$ by the  GDH experiment at MAMI. The small value
$\gamma_0(0)\approx[-1.01\pm0.08({\mbox{stat}})\pm0.10({\mbox{syst}})]
\cdot10^{-4}$~fm$^4$ is well reproduced by a strong cancellation of
S-wave pion production and $\Delta$ excitation in DR~\cite{Tia02}, while the
existing results of ChPT scatter around this value. The ChPT
prediction~\cite{Kao02}
at $\mathcal{O}(p^4)$, on the other hand, agrees quite well with the
result of DR, $\delta_{LT}(0)\approx1.4\cdot10^{-4}$~fm$^4$.
However, practically nothing is
known so far about $\delta_{LT}(Q^2)$ from the experimental side.
It will be interesting to study the $Q^2$ dependence of these
VVCS observables in more
detail, both theoretically and experimentally.

\end{document}